\pdfoutput=1

\documentclass{webofc}
\usepackage[varg]{txfonts}   % Web of Conferences font

\usepackage{amssymb,amsmath,amsfonts}
\usepackage[colorlinks,linkcolor=purple,citecolor=teal]{hyperref}
\usepackage{dsfont}
\usepackage{color}
\usepackage{graphicx}
\usepackage{hyphenat}
\usepackage{empheq}
\usepackage{textcomp}
\usepackage{rotating}
\usepackage{wrapfig}
\usepackage{pdfpages}
\usepackage{verbatim}
\usepackage{setspace}
\usepackage{array}
\usepackage{caption}
\usepackage[pdf]{pstricks}
\usepackage{upgreek}
\usepackage{cmll}
\usepackage{latexsym}
\usepackage{braket}

\usepackage{soul}
\sethlcolor{yellow}

\newcommand{\beq}{\begin{eqnarray}}
	\newcommand{\eeq}{\end{eqnarray}}
\newcommand{\bea}{\begin{eqnarray}}
	\newcommand{\eea}{\end{eqnarray}}
\newcommand{\be}{\begin{equation}}
	\newcommand{\ee}{\end{equation}}

\def\de{\partial}

\def\1{\mathbbm{1}}

\def\S{\mathcal{S}}

\def\Tr{\qopname\relax o{Tr}}

\numberwithin{equation}{section}

\newcommand{\bulk}{\Phi}
\begin{document}
\title{Neutron stars and phase diagram from a double hard-wall}
%
% subtitle is optionnal
%
%%%\subtitle{Do you have a subtitle?\\ If so, write it here}

\author{\firstname{Lorenzo} \lastname{Bartolini}\inst{1}\fnsep\thanks{Speaker, \email{lorenzo(at)henu.edu.cn}} \and
        \firstname{Sven Bjarke} \lastname{Gudnason}\inst{1}\fnsep\thanks{\email{gudnason(at)henu.edu.cn}} \and
        \firstname{Josef} \lastname{Leutgeb}\inst{2}\fnsep\thanks{\email{josef.leutgeb(at)tuwien.ac.at}} \and  
        \firstname{Anton} \lastname{Rebhan}\inst{2}\fnsep\thanks{\email{anton.rebhan(at)tuwien.ac.at}}
        % etc.
}

\institute{{Institute of Contemporary Mathematics, School of
	Mathematics and Statistics, Henan University, Kaifeng, Henan 475004,
	P.~R.~China}
\and
          {Institut
          für Theoretische Physik, Technische Universität Wien, Wiedner Hauptstrasse 8-10, A-1040 Vienna, Austria
      } 
          }

\abstract{%
 Description of nuclear matter in the core of neutron stars eludes the main tools of investigation of QCD, such as perturbation theory and the lattice formulation of the theory. Recently, the application of the holographic paradigm (both via top-down and bottom-up models) to this task has led to many encouraging results, both qualitatively and quantitatively. We present our approach to the description of neutron star cores, relying on a simple model of the (double) hard-wall type: we discuss results concerning the nature of homogeneous nuclear matter at high density emerging from the model including a quarkyonic phase, the mass-radius relation for neutron stars, as well as the rather stiff equation of state we have found.
 We show how, despite the very simple model employed, for an appropriate calibration we are able to obtain neutron stars that only slightly fall short of the observational bounds on radius and tidal deformability.
}
\maketitle
\section{Introduction}
\label{intro}
Determining the phase diagram of QCD is one of the most important issues in modern theoretical physics.
Unfortunately, ordinary matter is governed by the low-energy limit of QCD, where the theory runs into strong coupling and is hard to study theoretically.
Numerical computations of QCD in the lattice formulation are so far the most reliable sources of our knowledge about the phase diagram of QCD.
Unfortunately, lattice QCD works well only for  vanishing or small baryon chemical potentials, due to a technical problem known as the sign problem \cite{deForcrand:2009zkb,Aarts:2015tyj,Nagata:2021bru}.
In 1997 an alternative toolbox became available to theoretical physicists by the discovery of the AdS/CFT correspondence by Maldacena \cite{Maldacena:1997re}.
The exciting property about the duality is that the CFT at strong coupling is mapped to gravity at weak coupling, which means that perturbation theory in the bulk in a theory with one extra dimension can be used to study the field theory living on the boundary of AdS at strong coupling. 
A simple holographic setup was considered originally by Polchinski and Strassler where
the AdS space is cut off at a finite value of the radial coordinate
\cite{Polchinski:2001tt,Boschi-Filho:2002wdj,Polchinski:2002jw,Boschi-Filho:2002xih}.
Meson and baryons were then subsequently added to the model, providing the basis of perhaps the simplest possible phenomenologically viable holographic description of QCD \cite{deTeramond:2005su,Erlich:2005qh,DaRold:2005mxj,Hirn:2005nr,Karch:2006pv}, which
we shall take as basis for our study.
In order to have the possibility of separately adjusting the deconfinement and chiral restoration transitions in this model, we implement a so-called "double hard-wall" condition, where the wall for the gluons is kept at the hard-wall, but allowing for the flavor gauge and scalar fields to end on a second hard-wall. 
In this work we employ a homogeneous Ansatz, describing the instantons in the approximation suitable for large densities or equivalently finite/large chemical potential.
The phase diagram we find is rather rich phenomenologically, and is consistent with common lore for the different phases and their approximate placement in the diagram \cite{McLerran:2007qj,Fukushima:2013rx}.
We calculate the equation of state for nuclear matter, which for a range of densities of order of a few times the saturation density is a difficult region for other types of models to make predictions for.
The equation of state we find in our simple model is rather stiff compared to other holographic models in the literature \cite{Ishii:2019gta,Kovensky:2021kzl,BitaghsirFadafan:2019ofb}.
Though omitted in these proceedings, we also use the obtained equation of state to simulate a neutron star merger of two neutron stars of $1.4$ solar masses each at a separation distance of 45 kilometers and calculate the gravitational wave spectrum using a full-fledged numerical gravity and hydrodynamics code: for a detailed discussion of the simulation, as well as of all the other aspects of this work, see \cite{Bartolini:2022rkl}.
\section{The model}
\label{sec-1}

Largely following the setup and notation of \cite{Pomarol:2007kr,Domenech:2010aq},
the background geometry is taken to be that of a slice of AdS$_5$ ending at bulk coordinate $z_0$ with curvature scale indicated by $L$:
\beq
ds^2 = \frac{L^2}{z^2}\left(dx_\mu dx^{\mu} -dz^2\right).
\label{eq:metric}
\eeq
	The flavor field content is given by the presence of two $U(2)$ gauge vectors, $\mathcal{L}_M(x,z), \mathcal{R}_M(x,z)$ and a bi-fundamental complex scalar $\Phi(x,z)$ dual to the order parameter of chiral symmetry breaking. The action can be divided into three main contributions: a gauge part containing Yang-Mills-like terms, a Chern-Simons terms to account for flavor anomalies, and a piece containing kinetic and interaction terms for the scalar. The minimal action reads \cite{Pomarol:2007kr,Domenech:2010aq}:
\beq
S &=& S_g+S_{CS}+ S_{\Phi},\\\label{Sgauge}
S_{g}&=& -\frac{M_5}{2}\int d^4xdz\; a(z)\left[\Tr\left(L_{MN}L^{MN} \right)+\frac{1}{2}\widehat{L}_{MN}\widehat{L}^{MN} + \left\{R\leftrightarrow L\right\}\right],\\\label{SCS}
S_{CS}&=& \frac{N_c}{16\pi^2}\int d^4xdz\;\frac{1}{4}\epsilon_{MNOPQ}\widehat{L}_M\left\{\Tr\left[L_{NO}L_{PQ}\right]+\frac{1}{6}\widehat{L}_{NO}\widehat{L}_{PQ} -\left\{R\leftrightarrow L\right\}\right\},\\
S_{\Phi}&=& M_5 \int d^4xdz\; a^3(z)\left\{\Tr\left[\left(D_M\Phi\right)^\dag D^M\Phi\right] -a^2(z)M^2_{\bulk}\Tr \left[\Phi^\dag\Phi\right]\right\},
\eeq
where $a(z) = \frac{L}{z}$, 
$L_M=L_M^a\frac{\tau^a}{2}$ and $\widehat{L}_M$ are the $SU(2)$ and $U(1)$ parts of the $U(2)$ field $\mathcal{L}_M$, 
\beq
\mathcal{L}_M = L_M^a \frac{\tau^a}{2} + \widehat{L}_M\frac{\mathds{1}}{2}\qquad;\qquad D_M\Phi = \de_M \Phi -i\mathcal{L}_M\Phi +i \Phi \mathcal{R}_M,
\eeq
whose field strength is
$\mathcal{L}_{MN} = \de_{M}\mathcal{L}_{N} - \de_{N}\mathcal{L}_{M} -i\left[\mathcal{L}_M,\mathcal{L}_N\right]$; analogously $\mathcal{R}_{MN}$ is the field strength for the field $\mathcal{R}_M$, while $D_M\Phi$ is the covariant derivative of the scalar field.
For spacetime indices we use the following labels:
\beq
M,N,\ldots = {0,1,2,3,z},\qquad i,j,\ldots={1,2,3},\qquad \mu,\nu,\ldots ={0,1,2,3}.
\eeq

As shown in \cite{Herzog:2006ra}, the deconfinement transition happens via a Hawking-Page transition from the cutoff thermal AdS geometry to the AdS black hole geometry  described by the metric (not yet continued to Euclidean signature)
\beq
ds^2 = \frac{L^2}{z^2}\left(f(z)dt^2-dx_i^2-\frac{dz^2}{f(z)}  \right),\qquad f(z)= 1-\left(\frac{z}{z_h}\right)^4.
\eeq
The temperature of the dual theory is determined by the periodicity of the euclidean time coordinate $0 \le \tau < 1/T$, which is unconstrained in the thermal AdS case, but fixed in the black hole case by the regularity of the near-horizon solution as $T=(\pi z_h)^{-1}$.
For a critical value $z_c$ of the horizon position, and thus for a certain temperature $T_d$, the black hole phase becomes energetically favored, hence the deconfinement transition.
This transition depends only on the cutoff scale $z_{0}$ as $ T_d = 2^{1/4}(\pi z_0)^{-1}$.
Here $z_0$ is the cutoff for the geometry: we labelled it $z_{IR}$ before, as in usual hard-wall models the two concepts are identified and there is no ambiguity.

One issue with the usual hard wall approach is that the critical value $z_c$ of the black hole horizon location at which the phase transition happens is always lower than the cutoff $z_0=z_{IR}$.
The horizon of the black hole, as soon as it appears, hides the IR brane on which the spontaneous breaking of chiral symmetry is realized by $\xi$: this implies that as soon as the theory undergoes a deconfining transition, it also unavoidably restores chiral symmetry. 
However, in general these phase transitions can be widely separated \cite{Evans:2020ztq}.
To have a richer phase diagram with separate confined, deconfined, and chiral symmetry restored phases, we postulate that the geometrical cutoff and the gauge cutoff do not coincide, so we keep the notation $z_{IR}$ for the cutoff of the flavor gauge fields propagation which we permit to be smaller than $z_0$ where the geometry ends.

\section{Homogeneous baryonic matter}
\label{sec-2}

From the holographic point of view baryons are solitonic configurations in the flavor gauge fields: to build a many-instantons configuration accounting for interactions and minimization of the energy with respect to the moduli of the solitons is extremely challenging, so usually an approximation scheme is necessary to extract some qualitative result from the model. 
Another possibility is to take the fields to only depend on $z$ to begin with, instead of starting with single baryons and then performing some averaging.
A configuration like this would be far from reality when densities are low, but as the density increases the system tends to look homogeneous up to very short distances in $\mathds{R}^3$.
The resulting Ansatz is of the form: 
\begin{align}
	\label{hansatz}
	L_i = -R_i = -H(z)\frac{\tau^i}{2} \qquad ; \qquad \widehat{L}_0 = \widehat{R}_0 = \widehat{a}_0(z)\qquad;\qquad
	\Phi = \omega_0(z)\frac{\mathds{1}}{2},
\end{align}
with all other fields vanishing. The imposition of $R_i = -L_i$ is also necessary in order to obtain a nonvanishing baryon number. 

Given the Ansatz, we can extract a one-dimensional effective Lagrangian density from the model. To include the finite temperature theory we perform a Wick rotation as $t=-i\tau$. When the geometry undergoes a transition to the black hole background, the effective one-dimensional Lagrangian (and correspondingly the equations of motion) change, including additional blackening factors in some terms, so here we present the full effective Lagrangian in a notation that accounts for both phases, with $f(z)\to1$ for $T<T_d$:
\beq
\mathcal{L}^E_g 	&=&  M_5 a(z) \left[3H^4 + 3 f(z) H^{'2} - \widehat{a}_0^{'2}\right],\\
\mathcal{L}^E_{CS}&=&-\frac{3N_c}{8\pi^2}\widehat{a}_0 H^2H',\\
\mathcal{L}^E_\Phi &=& M_5a^3(z)\left[\frac{3}{2}H^2\omega_0^2 +\frac{1}{2}f(z) \omega'^2_0 +\frac{1}{2} M_{\bulk}^2 a^2(z) \omega_0^2\right],
\eeq
From this Lagrangian density we can obtain the grand-canonical potential via the holographic correspondence as $\Omega = -V \int dz\; \mathcal{L}.$

The boundary conditions to impose on the matter fields are related to quantities on the field theory side: in particular, the UV boundary condition on $\widehat{a}_0(z)$ is related to the baryon chemical potential, and the IR boundary conditions on $H(z)$ and $\omega_0(z)$ are related to the baryon density and the chiral condensate as follows:
\beq
H(z_{IR}) =  \left(4\pi^2 d\right)^{\frac{1}{3}}\label{eq:HIR},\qquad
\widehat{a}_0(z_{UV}) = \mu,\qquad
	\omega_0(z_{IR}) = \xi.
\eeq
Following \cite{DaRold:2005vr} the parameter $\xi$ can be thought of as originating from a variational principle, by minimizing the energy with the addition of an IR localized boundary term:
\beq\label{eq:SIR}
\S_{IR} = \sqrt{f(z_{IR})} \frac{m_b^2}{2} \xi^2 -  \lambda \xi^4.
\eeq
\section{Phase Diagram}
\label{sec-3}
\begin{figure}[h]
	\centering
	\includegraphics[width=0.49\linewidth]{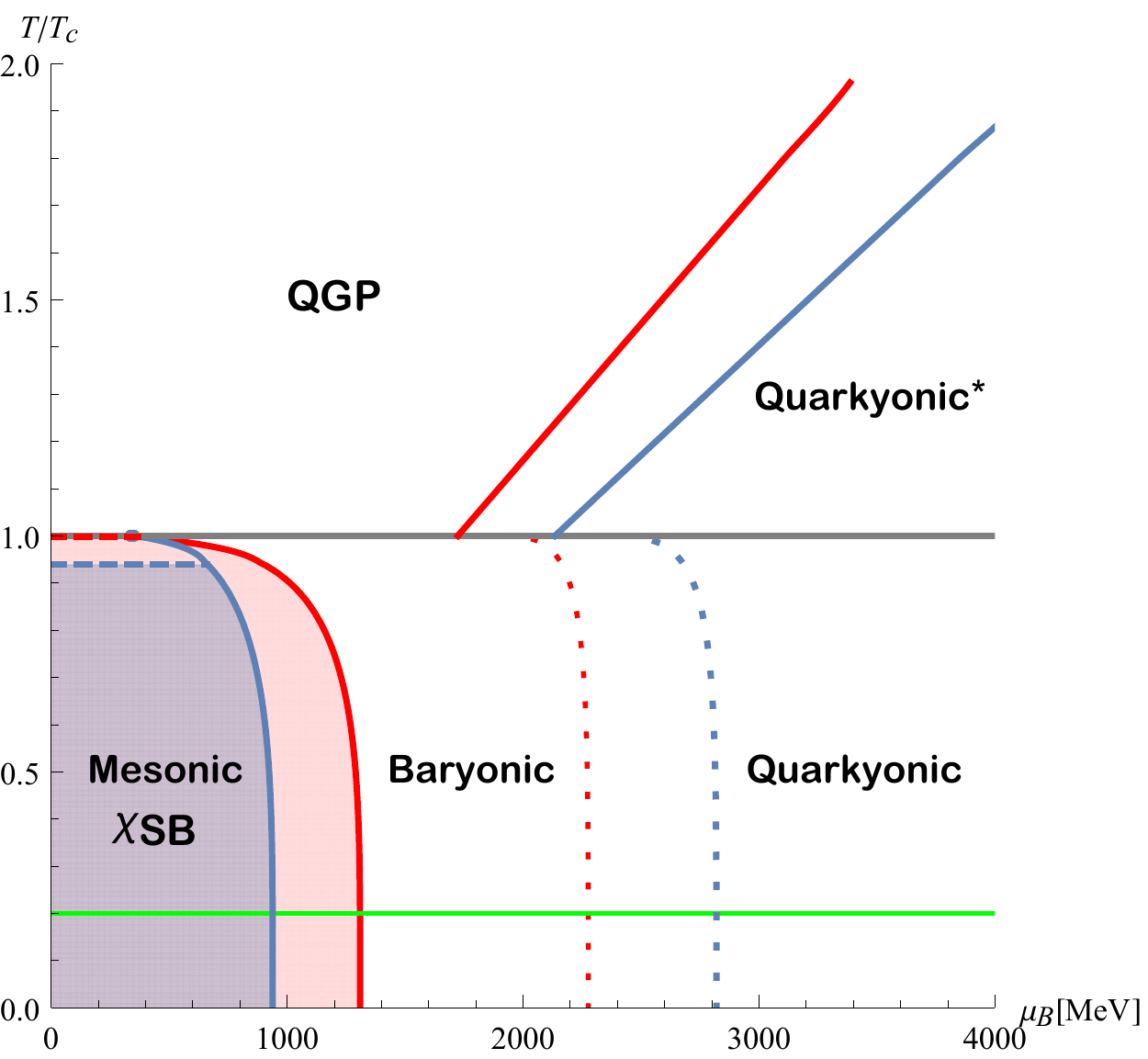}
	\includegraphics[width=0.49\linewidth]{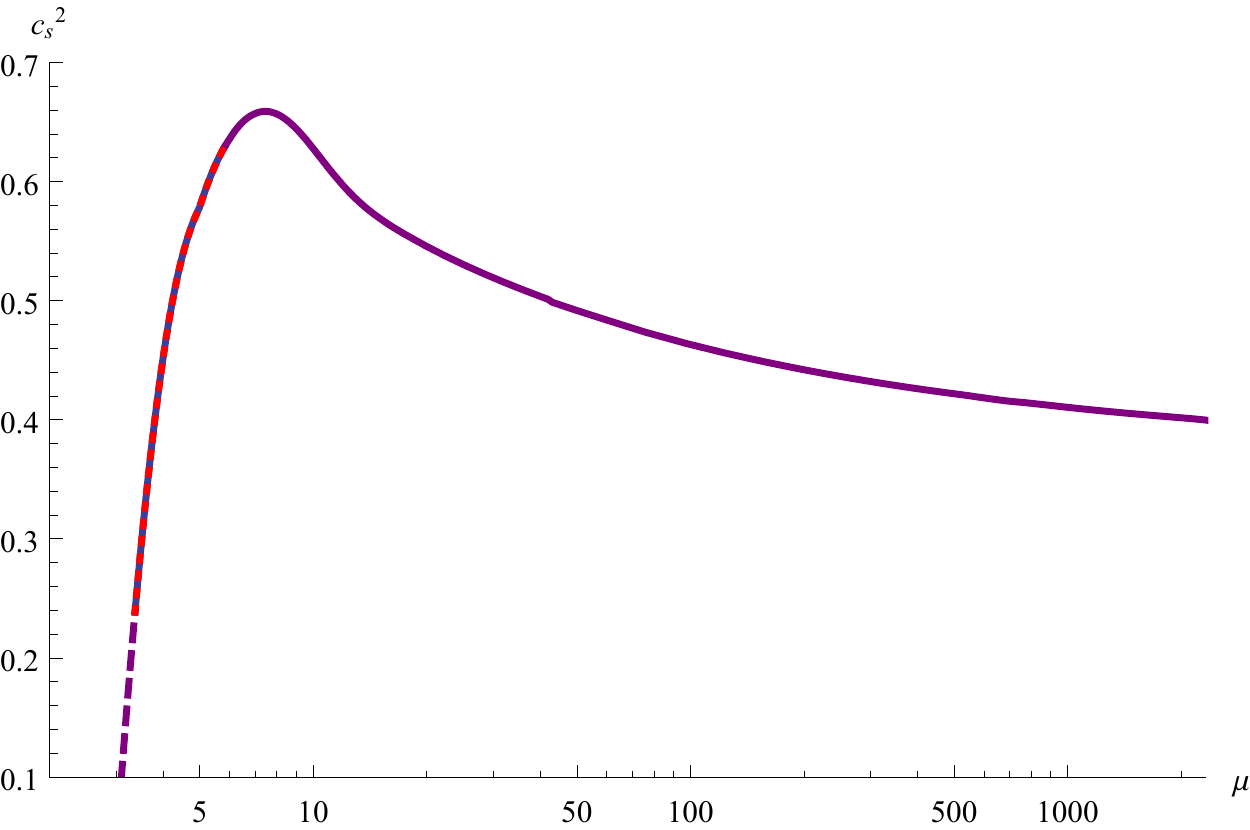}
	\caption{Left: Phase diagrams of holographic QCD from the double hard-wall model with blue and red colors representing respectively the choices of "Fit A" or "Fit B" (the purple lines representing overlap in lines of the two choices). The green line represents a(n arbitrary in our setup) deconfining transition, below which every transition line is perfectly vertical (there is no temperature dependence).
		The solid lines represent the corresponding first order transitions, while the dotted lines mark crossovers.
		The horizontal dashed red and blue lines represent a second-order transition to chiral restoration within the mesonic phase, with the colored regions identifying phases with broken chiral symmetry. Above the critical temperature, a phase of quark-gluon plasma exists, while for large enough chemical potentials, and as a function of temperature, baryons also appear in this phase, realizing a quarkyonic phase of a fundamentally different nature than the one below $T_c$ with coexistence of deconfined quarks and baryons. Right: Squared speed of sound in the baryonic phase as a function of the chemical potential $\mu$ in units where $L=1$. The dashed sections represent unstable branches, never realized since at those chemical potentials the mesonic phase is energetically favored.
		}
	\label{fig:Phasediag1}
\end{figure}
Here we analyze the phase diagram of the model. There are in principle a large variety of phases related to turning on and off different parameters: the chiral condensate $\xi$, the baryon density $d$, and also the quark density $k$ that we will see arising in the deconfined phase. All these quantities are (directly or indirectly) functions of the temperature $T$ and the chemical potential $\mu$: in fact the presence of couplings between the scalar field and $H(z)$ implies that a finite density (whose value depends on $\mu$) can indeed introduce corrections to the stable value of $\xi$. However the resulting picture is not as complicated as it could be, since the free energy has local minima only along the directions $d=0$ and $\xi=0$: this statement holds true within all the range of $T,\mu$ analyzed numerically, so that the search for the favored phase translates into the search for the global minimum between two local minima for every $\mu,T$. Moreover, for every $T<T_d$, the system shows no $T$-dependence.

To be able to draw the phase diagrams we need to fix some constants of the model. The two sets of choices for the free parameters we employed here (and that we will employ in the next section) are the following:

\beq\label{fitA}
\text{Fit A: } && L^{-1}=186{\rm MeV}\qquad ;\qquad \lambda =  2\times 10^{-3}\qquad ;\qquad \xi_0= 1.05,\\
\label{fitB}
\text{Fit B: }	&& L^{-1}=150{\rm MeV}\qquad ;\qquad \lambda \xi_0^4 = 1.024,
\eeq
where the fit A indicates one that correctly reproduces the baryon mass and the critical chemical potential for the baryon onset, and the fit B is chosen as an example of a fit that produces an equation of state which lies as much as possible within the constraints from observational data on neutron stars. For fit A the values of $\xi_0$ and $\lambda$ are independently relevant, while for fit B only the value $\lambda \xi_0^4$ enters the calculations.

We introduce temperature dependence by employing the deconfining geometry of the AdS black hole. Naively chiral symmetry is automatically restored as soon as the horizon of the black hole reaches $z_h=z_{IR}=1$ and hides the boundary conditions for the scalar field $\Phi(z_{IR})=\xi \mathds{1}$. 
However, before this happens, the true minimum of the energy can in principle be realized for $\xi=0$ even at lower temperatures, depending on the $\mu$ and $T$ dependence of the energy density and because of the $f(z)$ factor in (\ref{eq:SIR}). 
We find that this is indeed the case, and as previously mentioned the chiral restoration transition coincides with a baryonic onset. However, if $T>T_d$, the critical chemical potential for the baryon onset $\mu_{on}(T)$ is a strictly decreasing function of the temperature. Interestingly, for $T=T_c$, the curve ends at a finite value of $\mu_{on}(T_c)$: this way our phase diagram shows a triple point as expected for large-$N_c$ QCD.
However, depending on the choice of the parameters, the equilibrium value of $\xi$ reaches zero for temperatures lower than $T_c$ in a second order phase transition.

We also find a crossover to another phase: as we increase the density, we observe a continuous deformation of the baryon number density distribution in the holographic direction, as it changes from being peaked on the infrared brane to a configuration where it develops a second peak at a finite distance from the hard-wall.
It was argued that the distribution in the holographic direction should be related to a spectrum of energies for the condensed baryons \cite{Rozali:2007rx} and that this transition then indicates the onset of a quarkyonic phase for cold and dense nuclear matter \cite{Kaplunovsky:2012gb}. The nuclear homogeneous matter in our model exhibits the same feature, performing a (continuous, hence this transition being a crossover) ``popcorn transition''.
After this analysis, we can draw (fig. \ref{fig:Phasediag1}) the following phase diagram for our holographic QCD model: see the caption for details on the order of the transitions and the identification of the parameter sets used. Moreover, we present the result for the speed of sound, which develops a peak value as high as roughly $0.66$, signaling a rather stiff equation of state.
\section{Neutron Stars}
\label{sec-4}
 Unlike the holographic studies reviewed in \cite{Jarvinen:2021jbd,Hoyos:2021uff} we refrain from matching to realistic
equation of states for conventional nuclear matter at moderate densities, thereby completely neglecting any effects from the crust of a neutron star. 
Neutron stars are described by the Tolman-Oppenheimer-Volkov (TOV) equations, which read:
\beq
\frac{dP}{dr}=-G(\varepsilon+ P)\frac{m+4\pi r^3 P}{r(r-2Gm)}\qquad,\qquad\frac{dm}{dr}=4\pi r^2 \varepsilon.
\eeq
The equations are solved by using as a boundary condition $P(r=0)=P_0$ for a range of values of $P_0$, and the radius of the neutron star obtained is defined as the value of $R$ for which $P(R)=0$. Different values of $P_0$ will produce results of $R,m$ that define a curve in the $m$-$R$ plot to be compared with observational data.
The holographic dictionary identifies the grand potential with (minus) the on-shell action:
\beq
PV = -\Omega  = V \int dz \mathcal{L}^{\rm on-shell}\qquad ,\qquad \varepsilon = \frac{E}{V} = -P + \mu \frac{\partial P}{\partial \mu} = - P + N_c \frac{\mu}{2}d.
\eeq
\begin{figure}[h]
	\centering
	\includegraphics[width=0.52\linewidth]{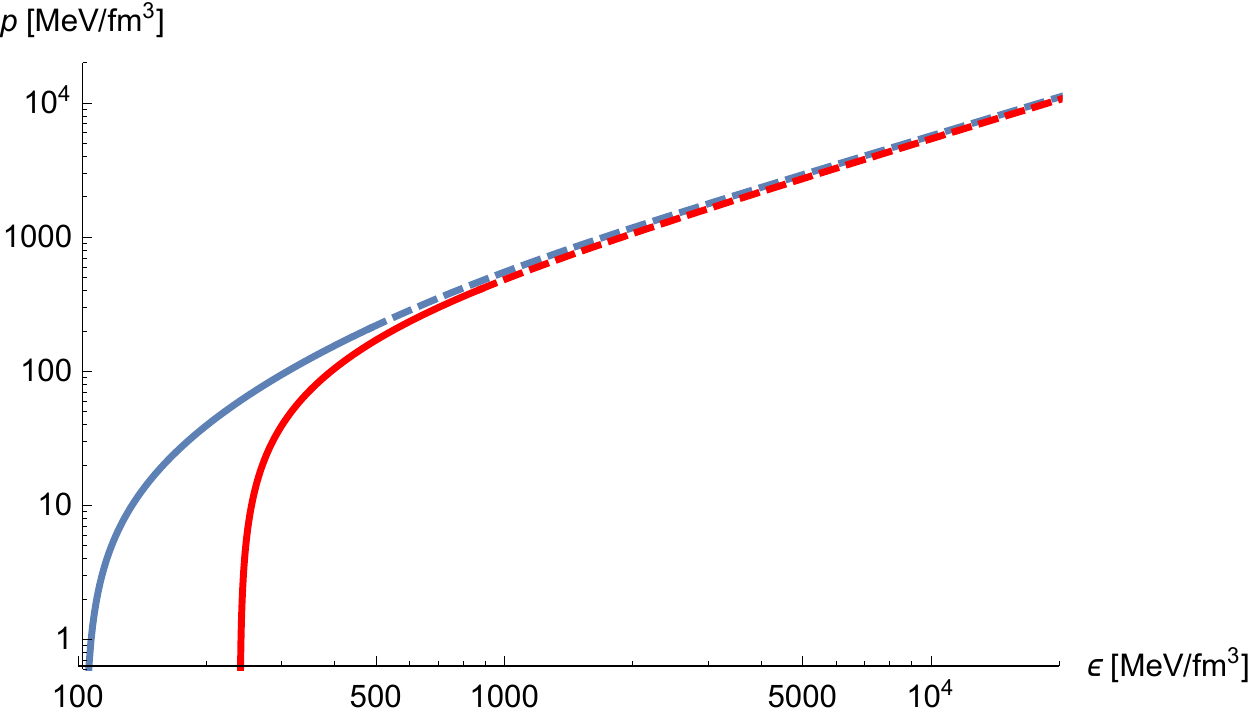}
	\includegraphics[width=0.47\linewidth]{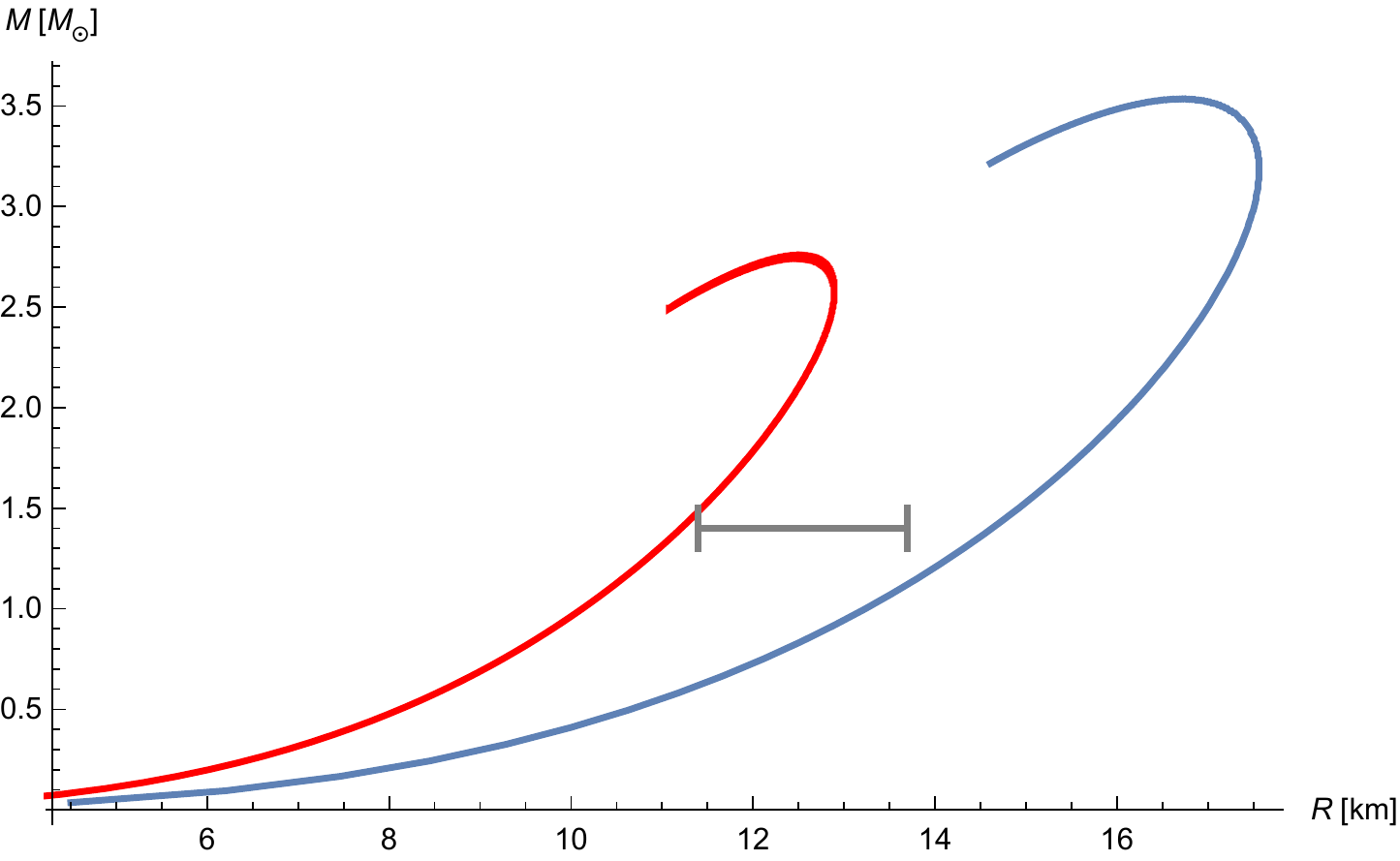}
	\caption{Equations of state (left) and radius-mass relations for neutron stars with the usual two choices of parameters (same color coding as previous figure). The dashed sections of the equations of state indicate values of central pressure that lead to unstable neutron stars. In the mass-radius plot, these stars lie beyond the turning point (corresponding to the maximum mass) of the respective curves.  }
	\label{eosMRplots}
\end{figure}
In fig.\ref{eosMRplots} we show the equation of state of our holographic nuclear matter, together with the characteristic $m-R$ curve of neutron stars built from it. The gray interval represents observational bounds for stars of mass $1.4M_\odot$. While "Fit A" leads to unrealistic stars, for "Fit B" all parameters are instead chosen to obtain a good compromise between highest mass, radius for stars of $1.4M_\odot$, and tidal deformability (see Fig.8 of \cite{Bartolini:2022rkl}). We see that in both cases the measured radius is not completely compatible with our predictions: however, the effect of a crust is expected to increase the radius of neutron stars, refining the precision of the ``phenomenological'' set of parameters. Moreover, in \cite{Bartolini:2022gdf} it is argued that proton fractions obtained with the same homogeneous Ansatz approach tend to be in the ballpark of the correct order of magnitude in this model at least for densities around saturation. 
\section{Conclusions}
\label{sec-5}

We studied the phase diagram of a holographic ``hard-wall'' model of QCD  with a nontrivial temperature dependence arising from a generalization to separate infrared walls for gluonic and quark degrees of freedom ("double hard-wall").
In the low-temperature phases, we have evaluated the resulting equation of state
with a view towards modelling cold nuclear matter at high densities.
Independently of the fit chosen for the free parameters in our model, the resulting speed of sound at zero temperature in this baryonic matter turned out to be rather high before monotonically decreasing with further increases of the chemical potential.

We used the equation of state of the model to solve the TOV equations and build neutron stars: the resulting neutron stars turn out to be quite compact, with maximum masses higher than what is expected, with the radius for stars of mass $1.4 M_\odot$ being slightly smaller than the bound, and with tidal deformability for the same star exactly on the lower bound allowed. The effects of the presence of a crust are expected to be an increase in the radius of stars (at fixed mass), and an increase in the tidal deformability:  this would open up the possibility to obtain a better phenomenological fit in order to obtain a lower maximum mass. Should the symmetry energy (and thus the proton fraction) also be lowered in this process, we would obtain a significant refinement towards realistic neutron stars completely built within holography (as done in \cite{Kovensky:2021kzl} with a top-down approach).

\bibliography{references}

\end{document}